\documentstyle[11pt,epsfig,twocolumn]{article}
\title{{\bfseries Critical curves of plane Poiseuille 
flow with slip boundary conditions}}
\author{{\bfseries Andreas Spille},
{\bfseries Alexander Rauh, and} 
{\bfseries Heiko B\"{u}hring}\\
{\small Carl von Ossietzky University Oldenburg, D-26111 Oldenburg, Germany}\\
{\small andreas.spille@gmx.de}}
\date{}

\begin{document}
\maketitle

\begin{abstract}
We investigate the linear stability
of plane Poiseuille flow in 2D under slip boundary conditions.
The slip $s$ is defined by the tangential velocity at the wall in 
units of the maximal flow velocity. As it turns out, the critical
Reynolds number depends smoothly on $s$ but increases quite
rapidly.
\end{abstract}

\section{Introduction}
No-slip boundary conditions are a convenient idealization of the behavior of
viscous fluids near walls. In real systems there is always a certain
amount of slip which, however, is hard to detect experimentally
because of the required space resolution. In high precision
measurements Elrick and Emrich [\ref{elrick}] detected slip of the order 0.1\%
in laminar pipe flow with Reynolds numbers of 16 to 4300.
The measuring error in [\ref{elrick}] was nearly as low as the 
fluctuations due to Brownian motion.
Very recently, Archer et al. [\ref{archer}] observed the existence of
slip in plane laminar Couette flow with added polymers.

We examine how the linear instability of the steady 
plane Poiseuille flow depends on the slip $s$ defined by
\begin{equation}
   s:=\frac{u_{\mbox{\tiny wall}}}{u_{\mbox{\tiny max}}}
\end{equation}
where $u_{\mbox{\tiny wall}}$ is the tangential velocity at the wall and
$u_{\mbox{\tiny max}}$ is the midstream velocity. As boundary conditions
we adopt
\begin{equation}
   \frac{\partial u}{\partial z}\pm b u=0,\quad w=0,\quad\mbox{at }z=\pm 1
   \label{randslip}
\end{equation}
where $z$ is measured in units of  the channel
half-width. The slip $s$ is implicitly determined by the parameter $b>0$
with $s\to 0$ in the limit $b\to\infty$.

\section{Orr-Sommerfeld equation with slip boundary conditions}

\begin{figure}[hbt]
\begin{center}
\begin{picture}(0,0)%
\epsfig{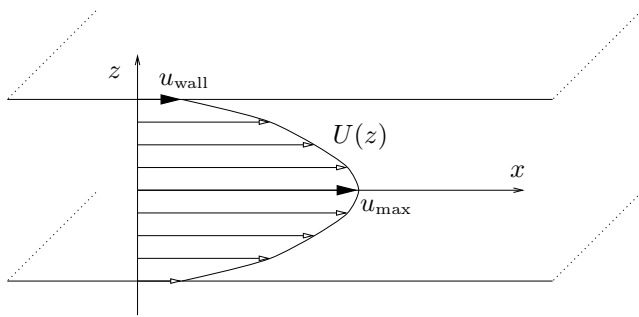}%
\end{picture}%
\setlength{\unitlength}{0.000400in}%
\begingroup\makeatletter\ifx\SetFigFont\undefined
% extract first six characters in \fmtname
\def\x#1#2#3#4#5#6#7\relax{\def\x{#1#2#3#4#5#6}}%
\expandafter\x\fmtname xxxxxx\relax \def\y{splain}%
\ifx\x\y   % LaTeX or SliTeX?
\gdef\SetFigFont#1#2#3{%
  \ifnum #1<17\tiny\else \ifnum #1<20\small\else
  \ifnum #1<24\normalsize\else \ifnum #1<29\large\else
  \ifnum #1<34\Large\else \ifnum #1<41\LARGE\else
     \huge\fi\fi\fi\fi\fi\fi
  \csname #3\endcsname}%
\else
\gdef\SetFigFont#1#2#3{\begingroup
  \count@#1\relax \ifnum 25<\count@\count@25\fi
  \def\x{\endgroup\@setsize\SetFigFont{#2pt}}%
  \expandafter\x
    \csname \romannumeral\the\count@ pt\expandafter\endcsname
    \csname @\romannumeral\the\count@ pt\endcsname
  \csname #3\endcsname}%
\fi
\fi\endgroup
\begin{picture}(8424,4074)(2389,-4423)
\put(7066,-3031){\makebox(0,0)[lb]{\smash{\SetFigFont{10}{12.0}{rm}$u_{\mbox{\tiny max}}$}}}
\put(3736,-1276){\makebox(0,0)[lb]{\smash{\SetFigFont{10}{12.0}{rm}$z$}}}
\put(4411,-1366){\makebox(0,0)[lb]{\smash{\SetFigFont{10}{12.0}{rm}$u_{\mbox{\tiny wall}}$}}}
\put(6706,-2131){\makebox(0,0)[lb]{\smash{\SetFigFont{10}{12.0}{rm}$U(z)$}}}
\put(9046,-2581){\makebox(0,0)[lb]{\smash{\SetFigFont{10}{12.0}{rm}$x$}}}
\end{picture}
\end{center}
\caption{Geometry of the basic flow with slip boundary conditions.}
\label{b_geometrie}
\end{figure}

The continuity equation in two dimensions is most conveniently
satisfied by introducing a stream function $\Psi(x,z,t)$ where
$x$ denotes the streamwise direction and $z$ the direction normal to
the boundaries (see Fig.\ \ref{b_geometrie}).
The velocity $(u,w)$ is connected to $\Psi$ through
\begin{equation}
   u=\frac{\partial \Psi}{\partial z},\quad
   w=-\frac{\partial \Psi}{\partial x}.
\end{equation}
In terms of $\Psi$ the Navier-Stokes equations for plane 
Poiseuille flow in two dimensions read in dimensionless
form
\begin{equation}
  \frac{\partial}{\partial t}\Delta\Psi
  +\frac{\partial\Psi}{\partial z}\frac{\partial\Delta\Psi}{\partial x} 
  -\frac{\partial\Psi}{\partial x}\frac{\partial\Delta\Psi}{\partial z}
  =\frac{1}{R}\Delta^2\Psi.
  \label{psi}
\end{equation}
As usually,
$\Psi$ is decomposed in the stream function $\Psi_b$ of the steady 
profile and a Fourier ansatz in $x$-direction for the disturbance field
with the wave number $\alpha$:
\begin{equation}
   \Psi(x,z,t)=\Psi_b(z)+\sum_{q=-\infty}^\infty e^{iq\alpha x}
    \Psi_q(z,t).
\end{equation}
However, with slip the basic flow is now given by
\begin{equation}
  \Psi_b=z-\frac{bs}{6}z^3;\quad s=\frac{2}{2+b}.
  \label{ugrund}
\end{equation}
The linearized part of (\ref{psi}) leads to the Orr-Sommerfeld equation 
\begin{equation}
   L\Psi_q=R\frac{\partial}{\partial t}(D^2-q^2\alpha^2)\Psi_q
\label{g1}
\end{equation}
where
\begin{equation}
   L=(D^2-q^2\alpha^2)^2-i\alpha qR[U(z)(D^2-q^2\alpha^2)-U''(z)]
\end{equation}
with $U(z)=\partial\Psi_b/\partial z$
and $D:=\frac{\partial}{\partial z}$. 

\section{Numerical method}

We determine the critical (neutral) curves in the parameter space
of the Reynolds number $R$ and the wave number $\alpha$ of the disturbance.
The Reynolds number is based on the channel half-width and
on the midstream velocity of the steady flow.

The solution of the differential equation (\ref{g1}) 
leads to a generalized eigenvalue problem
that we solve numerically as in [\ref{rzz}] using
up to 70 Chebyshev polynomials as basis functions.
The critical curve is the set of points $(R,\alpha)$ for
which the most critical eigenvalue has zero real part
with all other modes decaying exponentially.

\section{Results}

\begin{figure}[ht]
\vspace*{-2.5cm}
\begin{center}
\epsfig{figure=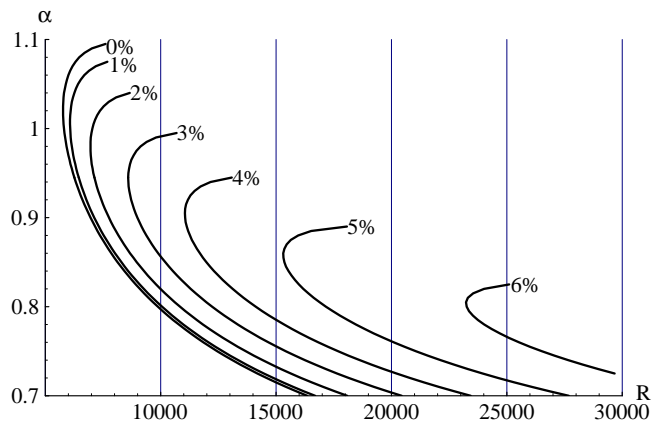,width=8.5cm}
\end{center}
\vspace*{-2cm}
\caption{Critical curves of plane Poiseuille flow for different slips
$s=0\%, 1\%, 2\%, 3\%, 4\%, 5\%$ and $6\%$.}
\label{b_slip}
\end{figure}

In Fig.\ \ref{b_slip} we present the critical
curves for different slips $s$.
The critical Reynolds number $R_c$ is the lowest Reynolds number
on the critical curve. We define also the slip $s_c$ 
by the corresponding normalized tangential velocity of the critical
mode at the wall.
The results are listed in Tab.\ \ref{t_slip}.

Obviously, the critical Reynolds number depends continuously on $s$.
However, there is, perhaps surprisingly, a strong increase both of 
$R_c$ and $s_c$ with increasing slip $s$.
In the limit $b\to\infty$, i.e. $s\to 0$, one gets the
well-known value $R_c\approx 5772$.

\begin{table}[ht]
%\begin{center}
\begin{tabular}{l|r|r|r|r|r}
$s$ & 0\% & 0.1\% & 0.2\% & 0.5\% & 1\% \\
\hline 
$s_c$ & 0\% & 0.9\% & 1.8\% & 4.5\% & 8\% \\
\hline
$R_c$ & 5772 & 5773 & 5781 & 5847 & 6070 
\end{tabular}
\begin{tabular}{l|r|r|r|r|r}
$s$ &2\%&3\%&4\%&5\%&6\%\\
\hline 
$s_c$ & 21\% & 31\% & 39\% & 47\% & 55\% \\
\hline
$R_c$ & 6960 & 8600 & 11060 & 15310 & 23230
\end{tabular}
%\end{center}
\caption{Slip $s_c$ of the critical mode and critical Reynolds number $R_c$
at different slips $s$ of the steady flow.}
\label{t_slip}
\end{table}
  
\section{Literature}

\renewcommand{\labelenumi}{[\theenumi]}
\begin{enumerate}
   \item \label{elrick} Elrick R.M., Emrich R.J., Phys. Fluids \textbf{9}
   (1966), 28 
   \item \label{archer} Archer L.A., Larson R.G., Chen Y.-L., 
   J.Fluid Mech. \textbf{301} (1995), 133
   \item \label{rzz} Rauh A., Zachrau T., Zoller J., Physica D {\bfseries 86} 
   (1995), 603
\end{enumerate}

\end{document}